\documentclass[12pt]{article}
\usepackage[centertags]{amsmath}
\usepackage{amsfonts}
\usepackage{amssymb}
\usepackage{amsthm} 
\usepackage{newlfont}
\usepackage{epsfig}
\usepackage{amscd}
\usepackage{graphicx}
\usepackage{epsfig}
\usepackage{showlabels}
\usepackage{color}
\usepackage{amscd}

\newcommand{\beq}{\begin{equation}}
\newcommand{\eeq}{\end{equation}}
\newcommand{\ba}{\begin{array}}
\newcommand{\ea}{\end{array}}
\newcommand{\bea}{\begin{eqnarray}}
\newcommand{\eea}{\end{eqnarray}}
\begin{document}

\begin{center}

{\large \sc \bf  {Unitary invariant discord as a measure of  bipartite  quantum correlations in an $N$-qubit quantum  system.
}}

\vskip 15pt

{\large 
 A.I.~Zenchuk 
}

\vskip 8pt

{\it Institute of Problems of Chemical Physics, Russian Academy of Sciences,
Chernogolovka, Moscow reg., 142432, Russia, e-mail:   zenchuk@itp.ac.ru
 } 
\end{center}

\begin{abstract}
We  introduce a measure of quantum correlations in the  $N$-qubit quantum system which is invariant with respect to the $SU(2^N)$ group of  transformations of this system. This measure is  a modification of the quantum discord   introduced earlier and is referred  to as the unitary or $SU(2^N)$-invariant discord.  Since the evolution of a quantum system is  equivalent to the proper unitary transformation, the introduced measure is an integral of motion and is completely defined by  eigenvalues of the density matrix. As far as  the calculation of the unitary invariant discord is rather complicated computational problem, we propose its modification which may be found in a simpler way. The case $N=2$ is considered in details.
In particular,  it is shown that the modified $SU(4)$-invariant discord   reaches the maximum value for  a pure state. 
 {A geometric measure of the unitary invariant discord of an  $N$-qubit state is introduced and a simple   formula for this measure 
 is derived, which allows one to consider this measure as a witness of quantum correlations.}
The relation of the unitary invariant discord with the quantum state transfer along the spin chain is considered.
We also compare the  modified $SU(4)$-invariant  discord  with the geometric measure of $SU(4)$-invariant  discord of the 
two-qubit systems in the thermal equilibrium states governed by the different Hamiltonians.
\end{abstract}

\section{Introduction}
The development of the quantum information technology stimulates a deep study of the properties of quantum correlations inherent in a quantum system. In particular, there is a problem of  identification of those quantum correlations which are responsible for the advantages of the quantum computations in comparison  with the classical ones. The entanglement \cite{W,HW,P,AFOV,HHHH}, which was originally taken as a measure of such correlations, seamed out to not cover all of them. {As a consequence, there are quantum systems without entanglement, which, nevertheless, either reveal a quantum nonlocality \cite{BDFMRSSW,HHHOSSS,NC} or }  speed-up certain calculations in comparison with the classical analogues \cite{M,DFC,DV,DSC,LBAW}. 
 Such observations cause a new stimulus  for  study those   quantum correlations which are not captured by entanglemet. Thus, the concept of  quantum discord is intensively developing diring last years \cite{OZ,L,ARA}.  Originally the quantum discord was introduced to characterize the impact of  the classical 
measurements on a quantum system with the purpose to get the maximal information about this system with the minimal influence on  it \cite{Z}.
At first glance, the quantum discord seems to cover all quantum correlations. However, it was shown that the quantum discord may be either bigger or smaller then the entanglement \cite{L,LL,L2}. Thus, we may state that the quantum discord involves different quantum correlations then entanglement does, in general, which  causes doubts whether the discord captures all correlations. Moreover, the discord is not symmetrical with respect to the subsystem chosen for the projective measurements \cite{OZ}, which stimulates  the futher study of this ambiguity \cite{FZ} and suggests to introduce the discord under  two-side projective measurements \cite{X} (a symmetrical discord). 
In addition, both the entanglement and  discord of bipartite $N$-qubit system are not invariant with respect to $SU(2^N)$ transformations. On the one hand, this allows the evolution of  discord which is important if we are interested in the strong quantum correlations between the two chosen subsystems. Owing to the  discord evolution, two originally uncorrelated subsystems may evolve to strongly correlated ones.  On the other hand, $SU(2^N)$ transformation means nothing but  a transformation to a new basis, which may be written in terms of the states of some new "virtual" particles. The discord between the new virtual particles may  differ from the discord between the original ones. However, it is not clear   which discord   (either between the original particles or between the  virtual ones) prevails yielding  advantages of the quantum devices in comparison with their classical analogues. Of course, if one has to effect the first subsystem  by means of the second one, then namely 
 the quantum correlations between these two subsystems are important so that the usual discord may be a proper measure of these correlations. But if one considers a quantum process inside of the quantum system, then the quantum correlations between the above mentioned "virtual" particles might be more important resulting to advantages of quantum devices.

{
Another motivation for the modification of the measure of quantum correlations is a revealing such quantum systems which have no quantumness. Let us remember that not any quantum system exhibits quantum correlations measured in terms of either entanglement or discord. Moreover, the family of quantum systems possessing quantum correlations is still not completely characterized. Thus, considering  the entanglement as a proper measure of quantum correlations in a  bipartite system (a system composed by two subsystems $A$ and $B$) \cite{W}, it was demonstrated that  separable  states, i.e.  states 
  representable in the form
\begin{eqnarray}\label{sep}
\rho_{A,B}=\sum_{i} p_i \rho^{(i)}_{A}\otimes \rho^{(i)}_{B},
p_i\ge 0,\;\;\sum_i p_i=1,
\end{eqnarray}
have zero entanglement and, consequently, may not be considered as candidates for implementation in quantum devices. 
Here 
$\rho^{(i)}_{A}$ and $\rho^{(i)}_{B}$ represent the density matrices in the Hilbert spaces of   
respectively subsystems $A$ and $B$.
 However, the discord as an alternative measure of quantum correlations \cite{OZ} may be nonzero even for  separable states (\ref{sep}). It was shown \cite{OZ,X} that the
discord under one-side projective measurements on subsystem $B$ is zero if $\rho_{A,B}$ is representable in the following form
\begin{eqnarray}\label{sep2}
\rho_{A,B} = \sum_i p_i \rho^{(i)}_A |b_i\rangle \langle b_i|, \;\;\sum_i p_i =1,
\end{eqnarray}
 where $\rho^{(i)}_A $, $i=1,2,\dots$, are density matrices in the Hilbert space of  the subsystem  $A$ and vectors $|b_i\rangle$, $i=1,2...$ represent an orthonormal basis in the Hilbert space of subsystem $B$.
For the  quantum discord under two-side projective measurements \cite{X}, it is shown  that  this discord  is zero only if a state is representable in the form 
\begin{eqnarray}\label{sep3}
\rho_{A,B}=\sum_{i,j} p_{ij} |a_i\rangle |b_j \rangle \langle a_i|\langle b_j|,\;\;p_{ij}>0,\;\;\sum_{i,j} p_{ij}=1,
\end{eqnarray}
where vectors $|a_i\rangle $, $i=1,2,\dots$, and $|b_j\rangle$, $j=1,2,\dots$, are some orthonormal bases in the Hilbert spaces of respectively subsystems $A$ and $B$. Both eqs.(\ref{sep2}) and (\ref{sep3}) deffer from eq.(\ref{sep}) in general.
Eq.(\ref{sep2}) means that 
  only  states diagonalizable  by local transformations (i.e. transformations in the Hilbert spaces of subsystems $A$ and $B$) have zero symmetrical quantum discord. Zero quantum discord means that all mutual information encoded in the system $A\cup B$ may be  revealed by classical measurements using a proper complete set of projective measurements. However, the fact that not any complete set of projective measurements may be used to reveal all mutual information may be considered as a quantumness of the system. In other words one can propose  that the system has no quantumness if only any complete  set of projective measurements exhibits all mutual information  encoded in the system $A\cup B$.
}

  All these  arguments  suggest us to modify (or generalize) the definition of the quantum  discord. Namely, let us introduce  a measure of quantum correlations which  takes into account the quantum correlations between all possible virtual spin-1/2 particles in the $N$-spin system. We refer to this measure as the unitary invariant discord and assume that namely this measure of quantum correlations  is responsible for the advantages of quantum computations in comparison with the classical ones. The family of states with zero unitary invariant discord is revealed in Sec.\ref{Section:zero_d}. {{ It is shown that the new measure may be zero only for quantum states whose density matrix is proportional to the unit one. }

It is seemed out that the calculation of  unitary invariant discord is a complicated computational problem involving both the multidimensional optimization and the multiple integration, see Sec.\ref{Section:SU4}. Therefore  we introduce a modification of the unitary invariant discord  which  requires only the multiple integration without optimization. In addition, a geometric measure \cite{HHH,WG,DVB,LF}  of the unitary invariant discord is proposed, which seemed out to coinside with the geometric measure of modified unitary invariant discord.
The feature of the introduced geometric measure is that  its calculation does not involve neither optimization nor multiple integration. It has a  simple analytical representation in terms of  eigenvalues of the density matrix. {  For this reason, the geometric measure may be taken as a witness of quantum correlations in a system. 
The simplicity of the geometric measure of unitary invariant discord is especially important because the calculation of usual descord  \cite{OZ} is a complicated procedure which has been carried out only for some particular states, see for instance \cite{L,ARA,X}}.

Hereafter, a system of $N$ spin-1/2 particles is referred to as  $N$-qubit system.
This paper is organized as follows. The unitary invariant discord for a  two spin-1/2 particle state  ($SU(4)$-invariant discord) is introduced and studied in Sec.\ref{Section:method}. A modification of the $SU(4)$-invariant  discord is suggested as a relatively simple method to estimate this measure of quantum correlations in Sec.\ref{Section:mod}. It is shown that this modification reaches its maximal value for a pure state. The concept of unitary invariant discord is generalized for a system of $N>2$ spin-1/2 particles  ($SU(2^N)$-invariant discord) in Sec.\ref{Section:Nparticles}. A geometric measure of the unitary invariant discord is proposed in Sec.\ref{Section:geometric}. Comparison of the modified $SU(4)$-invariant discord with the normalized  geometric measure of the $SU(4)$-invariant discord is given in Sec.\ref{Section:Therm} for the thermal equilibrium states with different Hamiltonians.
The basic results are collected in Sec.\ref{Section:conclusions}.


\section{A system of two spin-1/2 particles}
\label{Section:method}
Let us consider a system of two spin-1/2 particles $A$ and $B$ (a two-qubit system) in the standard multiplicative basis associated with the above two particles: 
\begin{eqnarray}\label{alpha}
|\alpha_1 \rangle = |00\rangle,\;\;
|\alpha_2 \rangle = |01\rangle,\;\;
|\alpha_3 \rangle = |10\rangle,\;\;
|\alpha_4 \rangle = |11\rangle.
\end{eqnarray}
Here 1 and 0 mean the spin  directed along  and opposite the fixed $z$-axis respectively.
Let us consider an arbitrary special unitary transformation $U(\varphi)\in SU(4)$, where  the set of independent parameters $\varphi = (\varphi_1,\dots,\varphi_{15})$ parametrizes the group  $SU(4)$ with $U(0)=I_4$ (hereafter $I_n$ is the $n\times n$ identity matrix). This transformation allows us to introduce another basis 
\begin{eqnarray}\label{basis}
 |\beta_i(\varphi)\rangle= \sum_{j=1}^4   U_{ij}^*(\varphi)|\alpha_j\rangle,\;\;
|\beta_i(0)\rangle\equiv |\alpha_i\rangle,\;\;\;i=1,\dots,4.
\end{eqnarray}
The basis $|\beta_i(\varphi)\rangle$ may be treated as  the multiplicative basis of two   virtual spin-1/2 particles $A^\varphi$ and $B^\varphi$. In other words, we may write vectors $|\beta_i(\varphi)\rangle$ in the  following form {similar to the representation (\ref{alpha})}: 
\begin{eqnarray}\label{beta}
|\beta_1(\varphi) \rangle = |00\rangle_\varphi,\;\;
|\beta_2(\varphi) \rangle = |01\rangle_\varphi,\;\;
|\beta_3(\varphi) \rangle = |10\rangle_\varphi,\;\;
|\beta_4(\varphi) \rangle = |11\rangle_\varphi.
\end{eqnarray}
Now the density matrix  $\rho$ for the above system may be written using either the  basis $|\alpha_i\rangle$ or the basis  
$|\beta_i(\varphi)\rangle $, $i=1,\dots,4$, as follows:
\begin{eqnarray}
\rho =\sum_{i,j=1}^4 \rho_{ij}(0) |\alpha_i\rangle \langle \alpha_j|=\sum_{i,j=1}^4 \rho_{ij}(\varphi)|\beta_i(\varphi)\rangle  \langle \beta_j(\varphi)|,\;\;\;|\beta_i(0)\rangle \equiv |\alpha_i\rangle.
\end{eqnarray} 
Introduce the matrices $\rho(\varphi)=\{\rho_{ij}(\varphi)\}$ and $U(\varphi)=\{U_{ij}(\varphi)\}$.
The transformation betwen bases $|\alpha_i\rangle$ and $|\beta_i(\varphi)\rangle$ given by eq.(\ref{basis}) yields the evident relation:
\begin{eqnarray}\label{varphi}
\rho(\varphi)=U(\varphi) \rho(0) U^+(\varphi).
\end{eqnarray}
Here $\rho(0)$ and $\rho(\varphi)$ are the density matrices written in the bases $|\alpha_i\rangle $ and $|\beta_i(\varphi)\rangle $ respectively and, consequently, these matrices describe the systems of  two particles $A\cup B$ and $A^\varphi \cup B^\varphi$ respectively.
Of course,  the discord between the particles $A$ and  $B$ differs from  the discord between  the (virtual) 
particles  $A^\varphi$ and  $B^\varphi$. However, as it is  mentioned in the Introduction,  both discords measure the quantum correlations in the same two-qubit system and it is impossible to predict which  discord is really responsible for the advantages of quantum devices in comparison with their classical counterparts. This is a motivation to introduce a measure of quantum correlations which takes into account correlations between all virtual particles  $A^\varphi$ and  $B^\varphi$.  

Remark that relation (\ref{varphi}) may be viewed as a  transformation of the matrix $\rho(0)$ itself rather then the  transformation of the basis. This means that
the  discord between two virtual particles $A^\varphi$ and $B^\varphi$  is equivalent to the discord between the original particles $A$ and $B$  whose state is transformed by the unitary transformation $U(\varphi)$: 
$U(\varphi)  \rho(0) U^+(\varphi)$. This evident remark will be used for the calculation of the discord between the virtual particles $A^\varphi$ and $B^\varphi$ hereafter.

\subsection{The SU(4) invariant discord and its modification}
\label{Section:SU4}
The quantum discord as a measure of quantum correlations in a bipartite system has been introduced in ref.\cite{OZ}. This definition is based on the projective measurements over one of the subsystems. Since there is an asymmetry with respect to the subsystem chosen for the  projective measurements \cite{OZ,FZ}, a two-side measurement generalization of the bipartite discord has been proposed in ref.\cite{X} where the above asymmetry disappears.
We use the discord under the two-side projective measurements hereafter.

Let us recall the formula defining the discord for the system consisting of two spin-1/2 particles $A^\varphi$ and 
$B^\varphi$ with the density matrix $\rho(\varphi)$ \cite{OZ,X}:
\begin{eqnarray}\label{QQ}
q_{AB}(\rho(\varphi)) = I(\rho(\varphi)) - \sup_{\{\Pi\}} I(\tilde \rho(\varphi)).
\end{eqnarray}
Here 
\begin{eqnarray}\label{cl}
I(\rho(\varphi))=S(\rho^{A}(\varphi)) +S(\rho^{B}(\varphi))-
 S(\rho(\varphi)) ,
\end{eqnarray}
and $\tilde \rho(\varphi)$ (corresponding to the two-side projective measurements \cite{X}) is defined as follows:
\begin{eqnarray} \label{trho2s}
&&
\tilde \rho(\varphi) = \sum_{i,j=1}^2 \Pi_{ij}(\varkappa) \rho(\varphi) \Pi_{ij}(\varkappa),
\end{eqnarray}
where  $\Pi_{ij}(\varkappa) $ have the following form:
\begin{eqnarray}
\label{Pilrs}
 \Pi_{ij}(\varkappa) &=&\Pi_{i }(\varkappa_A)  \otimes \Pi_{j}(\varkappa_B) ,\;\;\;\varkappa=(\varkappa_A,\varkappa_B),\;\;i,j=1,2
\end{eqnarray}
and
\begin{eqnarray}\label{Pi}
&&
\Pi_{i}(\varkappa_A) = V(\varkappa_A)\Pi_{i }(0) V^+(\varkappa_A),\;\;
\Pi_{i}(\varkappa_B) = V(\varkappa_B)\Pi_{i }(0) V^+(\varkappa_B),\\\nonumber
&&
 V(\varkappa_A),\;V(\varkappa_B)\in SU(2),\;\;\;\;
\Pi_{1}(0)=|0\rangle \langle 0 |,\;\;\Pi_{2}(0)=|1\rangle \langle 1|.
\end{eqnarray}
The vector parameters $\varkappa_A=\{\varkappa_{Ai},\;i=1,2,3\}$ and $\varkappa_B=\{\varkappa_{Bi},\;i=1,2,3\}$  represent two different parametrizations of the group $SU(2)$. Of course, these parameters depend on $\varphi$, i.e. $\varkappa_{Ai}=\varkappa_{Ai}(\varphi)$ and $\varkappa_{Bi}=\varkappa_{Bi}(\varphi)$.
The optimization in eq.(\ref{QQ}) yields the proper values for the parameters $\varkappa_{A,B}$, which, in turn, depend on $\varphi$.  quantumness
This optimization  provides the invariance of the discord  with respect to 
the local $SU(2)$ transformations of particles $A$ and $B$.
Next, in order to take into account the discords between all possible virtual particles $A^\varphi$ and $B^\varphi$ we  average $q_{AB}(\rho(\varphi))$ over $\varphi$. However, before proceed to calculations, let us recall that there is a freedom in the choice of  the initial basis (i.e. the basis corresponding to $\varphi=0$) and the result depends on this initial basis. To remove this ambiguity we suggest to 
take the basis of eigenvectors of $\rho$ arranged in the order of decreasing eigenvalues as the original basis for the subsequent calculations. 
The density matrix   $\rho$ is diagonal in this basis:
\begin{eqnarray}\label{Lam}
&&
\rho(0)=\Lambda_2={\mbox{diag}} \{\lambda_1,\lambda_2,\lambda_3,\lambda_4\},\;\;\;\lambda_4=1-\sum_{i=1}^3\lambda_i, \;\;
\\\nonumber
&&
0\le \lambda_i\le 1,\;\;\;\lambda_1\ge \lambda_2\ge \lambda_3\ge \lambda_4.
\end{eqnarray} 
Thus, $\lambda_1$ is the maximal eigenvalue hereafter.
Using eq.(\ref{Lam}), we rewrite eq.(\ref{varphi}) as
\begin{eqnarray}
\rho(\varphi)= U(\varphi) \Lambda_2 U^+(\varphi).
\end{eqnarray}
Now the averaged value  of $q_{AB}(\rho(\varphi))$ can be calculated as  follows:
\begin{eqnarray}\label{barQ}
&&
\bar q_{AB}(\Lambda_2)=\int d\Omega(\varphi) q_{AB}(\rho(\varphi))=\\\nonumber
&&
\int d\Omega(\varphi)I(\rho(\varphi)) -\sup_\varkappa \int d\Omega(\varphi) I\Big(\sum_{ij}\Pi_{ij}(\varkappa(\varphi)) \rho(\varphi)\Pi_{ij}(\varkappa(\varphi)) \Big),
\end{eqnarray} 
where $\Omega$ is a measure in the space of parameters $\varphi_i$,  $\int d\Omega(\varphi)=1$ {{ and $\Pi_{ij}$ are defined in eq.(\ref{Pilrs})}.
Remark, that the averaged discord $\bar q_{AB}$ may be simply calculated for a pure state when 
$q_{AB}(\rho(\varphi))=S(\rho^A(\varphi))=S(\rho^B(\varphi))$ \cite{DSC}. In this case, there is no optimization in 
eq.(\ref{barQ}) and one obtains $\bar q_{AB}(\Lambda^{pure}_2) \approx 0.42$, 
where 
\begin{eqnarray}
\label{Lam_2_pure}
\Lambda^{pure}_2={\mbox{diag}}\{1,0,0,0\}.
\end{eqnarray}

{{ 
Now we represent a list of important properties of $\bar q_{AB}(\Lambda_2)$.
\begin{enumerate}
\item
Due to the averaging, the quantity $\bar q_{AB}(\Lambda_2)$ is completely defined by  the eigenvalues of the matrix $\rho$.
\item
As a consequence of the previous property, 
$\bar q_{AB}(\Lambda_2)$ is invariant  with respect to $SU(4)$ transformations of the whole  system. 
\item
Since discord $q_{AB} \ge 0$, one has that $\bar q_{AB}(\Lambda_2) \ge 0 $ by its definition (\ref{barQ}). 
\item
The averaged discord  $\bar q_{AB}$ is an  integral of motion, i.e. it  does not evolve. In fact, any evolution is described by the Liouville equation $i \rho_t(t) = [H(t),\rho(t)]$ ($\hbar=1$) with some initial condition $\rho|_{t=0}=\rho_0$ and the Hamiltonian $H$. 
Thus, the evolution of the density matrix $\rho$ may be written as $\rho(t)=U(t) \rho_0 U^+(t)$, $U(t)\in SU(4)$. This means that the density matrix $\rho(t)$ has the same eigenvalues as $\rho_0$ and consequently the unitary invariant discord for $\rho(t)$ equals to that for  $\rho_0$.
\end{enumerate}
A consequence of the property 2 is that $\bar q_{AB}$ is invariant  with respect to the local  $SU(2)$ transformations as well, similar to the usual discord.
}

The averaged discord $\bar q_{AB}(\Lambda_2)$
 can be normalized as follows:
\begin{eqnarray}\label{Qnorm}
Q_{AB}(\Lambda_2)=\frac{\bar q_{AB}(\Lambda_2)}{ \bar q_{AB}(\Lambda_2^{pure})},
\end{eqnarray}
so that $Q_{AB}(\Lambda^{pure}_2)=1$. 
The quantity $Q_{AB}(\Lambda_2)$ is referred to as the unitary invariant discord hereafter. {{ It possesses the same properties 1 -- 4 as the averaged discord $\bar q_{AB}(\Lambda_2)$.}

\subsubsection{Modified unitary invariant discord}
\label{Section:mod}
The calculation of $ \bar q_{AB}(\Lambda_2)$ is a 
 complicated computational problem because the second integral in eq.(\ref{barQ}) involves an optimization over $\varkappa(\varphi)$.
Thus, it is reasonable to introduce another quantity which would be simpler for evaluation and gives a reasonable estimation of $ \bar q_{AB}(\Lambda_2)$. The upper bound of 
 $\bar q_{AB}$ seems  to be such quantity. This upper bound may be readily written due to the obvious inequality
\begin{eqnarray}
&&
\sup_\varkappa \int d\Omega(\varphi) I\Big(\sum_{ij}\Pi_{ij}(\varkappa(\varphi) \rho(\varphi)\Pi_{ij}(\varkappa(\varphi))\Big) \ge
 \int d\Omega(\varphi) I\Big(\sum_{ij}\Pi_{ij}(0) \rho(\varphi)\Pi_{ij}(0)\Big) ,
\end{eqnarray}
which suggests us to 
define the upper bound of $\bar q_{AB}$ as follows:
\begin{eqnarray}\label{barQup}
\bar q^{up}_{AB}(\Lambda_2)=
\int d\Omega(\varphi)I(\rho(\varphi)) -\int d\Omega(\varphi) I\Big(\rho^d(\varphi)\Big),\;\;\;\;
\rho^d(\varphi)=
\sum_{ij}\Pi_{ij}(0) \rho(\varphi)\Pi_{ij}(0),
\end{eqnarray} 
where $\rho^d(\varphi)$ is the diagonal part of $\rho(\varphi)$. Similar to $\bar q_{AB}$, its upper bound  $\bar q^{up}_{AB}$ depends only on the eigenvalues of the matrix $\rho(0)$.

One has to emphasize that the upper bound $\bar q^{up}_{AB}(\Lambda_2)$  may be viewed as a difference between 
the total (i.e. quantum and classical) mutual information encoded in the two-qubit system  and its classical contribution encoded in the diagonal part of the matrix $\rho(\varphi)$ for all $\varphi$. From this point of view, the value  ${{\bar q}}^{up}_{AB}(\Lambda_2)$
 may be considered as an alternative measure of quantum correlations. 
Then 
it is reasonable to normalize   $\bar q^{up}_{AB}$ as follows
\begin{eqnarray}\label{calQ}
{\cal{Q}}_{AB}(\Lambda_2)=\frac{\bar q^{up}_{AB}(\Lambda_2)}{
\bar q^{up}_{AB}(\Lambda_2^{pure})},
\end{eqnarray}
so that ${\cal{Q}}_{AB}(\Lambda_2^{pure})={{Q}}_{AB}(\Lambda_2^{pure})=1$.
By construction, the value ${\cal{Q}}_{AB}$  is not an upper bound of the unitary invariant  discord $Q_{AB}$ in general. We refer to ${\cal{Q}}_{AB}(\Lambda_2)$ as the modified unitary invariant discord.
Emphasize that both the normalization $\bar q_{AB}(\Lambda_2^{pure})$ in eq.(\ref{Qnorm}) and  the normalization 
$\bar q^{up}_{AB}(\Lambda_2^{pure})$ in eq.(\ref{calQ}) are the universal constants, i.e. they  remain the same for all two-qubit systems. 

It will be shown below in this section  that $ \bar q^{up}_{AB}(\Lambda_2^{pure})\approx 0.69 $ and ${\cal{Q}}_{AB}(\Lambda_2)\le 1$. {{ In addition, 
it is obvious, that the modified  unitary invariant discord possesses the same list of properties 1--- 4 as the unitary invariant discord.}

Finally remark that there is only one possible bipartite decomposition of the system of two spin-1/2 particles, so that 
we may write 
\begin{eqnarray}\label{QAB}
Q(\Lambda_2)\equiv Q_{AB}(\Lambda_2),\;\;\;{\cal{Q}}(\Lambda_2)\equiv {\cal{Q}}_{AB}(\Lambda_2).
\end{eqnarray}

\paragraph{Modified $SU(4)$-invariant discord as a function of  density matrix eigenvalues.}
\label{Section:ev}

 We investigate the $\Lambda_2$-dependence of the modified  unitary invariant discord ${\cal{Q}}(\Lambda_2)$ 
in this paragraph. This will give us a clue 
to reveal systems with strong quantum correlations. For this purpose, let us
use the parametrization of an arbitrary $U(\varphi)\in SU(4)$ proposed in ref. \cite{TBS}:
\begin{eqnarray}\label{U}
U(\varphi)&=&
e^{i\gamma_3 \varphi_1}e^{i\gamma_2 \varphi_2}e^{i\gamma_3 \varphi_3}e^{i\gamma_5 \varphi_4}
e^{i\gamma_3 \varphi_5}e^{i\gamma_{10} \varphi_6}e^{i\gamma_3 \varphi_7}\times 
\\\nonumber
&&
e^{i\gamma_2 \varphi_8}e^{i\gamma_3 \varphi_9}
e^{i\gamma_5 \varphi_{10}}e^{i\gamma_{3} \varphi_{11}}e^{i\gamma_2 \varphi_{12}}e^{i\gamma_3 \varphi_{13}}e^{i\gamma_8 \varphi_{14}}e^{i\gamma_{15} \varphi_{15}},
\end{eqnarray}
where $4\times 4$ matrices $\gamma_i$ are given in the Appendix, eqs.(\ref{gamma}).
We parametrize $\Lambda_2$ by three parameters $\lambda_i$, $i=1,2,3$, as it is shown in eq.(\ref{Lam}).
Eq.(\ref{Lam})  represents the most general structure of matrix $\Lambda_2$ with the eigenvalues $\lambda_i$ on the diagonal arranged in the decreasing order. Note that not all parameters $\varphi_i$ appear in the matrix $\rho(\varphi)=U(\varphi)\Lambda_2 U^+(\varphi)$. In fact, 
$\gamma_3$, $\gamma_8$ and $\gamma_{15}$ are diagonal matrices (see eqs.(\ref{gamma})) so that  they commute with each other and with $\Lambda_2$. Thus, 
the formula for $\rho(\varphi)$ reads:
\begin{eqnarray}\label{UrhoU}
&&
\rho(\varphi)= U(\varphi) \Lambda_2 U^+(\varphi)=\tilde U(\varphi) \Lambda_2 \tilde U^+(\varphi),\\\nonumber
&&
\tilde U(\varphi)=e^{i\gamma_3 \varphi_1}e^{i\gamma_2 \varphi_2}e^{i\gamma_3 \varphi_3}e^{i\gamma_5 \varphi_4}
e^{i\gamma_3 \varphi_5}e^{i\gamma_{10} \varphi_6}e^{i\gamma_3 \varphi_7}e^{i\gamma_2 \varphi_8}e^{i\gamma_3 \varphi_9}
e^{i\gamma_5 \varphi_{10}}e^{i\gamma_{3} \varphi_{11}}e^{i\gamma_2 \varphi_{12}},
\end{eqnarray}
which possesses 12 parameters $\varphi_i$, $i=1,\dots,12$.
Here ranges of parameters $\varphi_i$ are following \cite{TBS}
\begin{eqnarray}
&&
0\le \varphi_1,\varphi_3, \varphi_5,\varphi_7, \varphi_9,\varphi_{11} \le \pi,\;\;\;
0\le \varphi_2,\varphi_4,\varphi_6,\varphi_8,\varphi_{10},\varphi_{12} \le \frac{\pi}{2}
,
\end{eqnarray}
so that $\displaystyle d\Omega(\varphi)= \frac{2^{6}}{\pi^{12}} d\varphi_1\dots d\varphi_{12}$.

To analyze the $\Lambda_2$-dependence of the modified unitary invariant discord  ${\cal{Q}}$, we, first of all, consider the case of at most two nonzero eigenvalues $\lambda_i$: $\lambda_2=1-\lambda_1$, $
\lambda_3=\lambda_4=0$.
Thus, there is one independent parameter $\lambda_1$ in this case. The graph of the function ${\cal{Q}}(\Lambda_2)\equiv {\cal{Q}}(\lambda_1)$ is shown in Fig.\ref{Fig:12}$(a)$. Here ${\cal{Q}}\gtrsim 0.72$, i.e. the modified  $SU(4)$-invariant discord may not be less then $\approx 0.72$ if there are two zero eigenvalues.

\begin{figure*}
   \epsfig{file=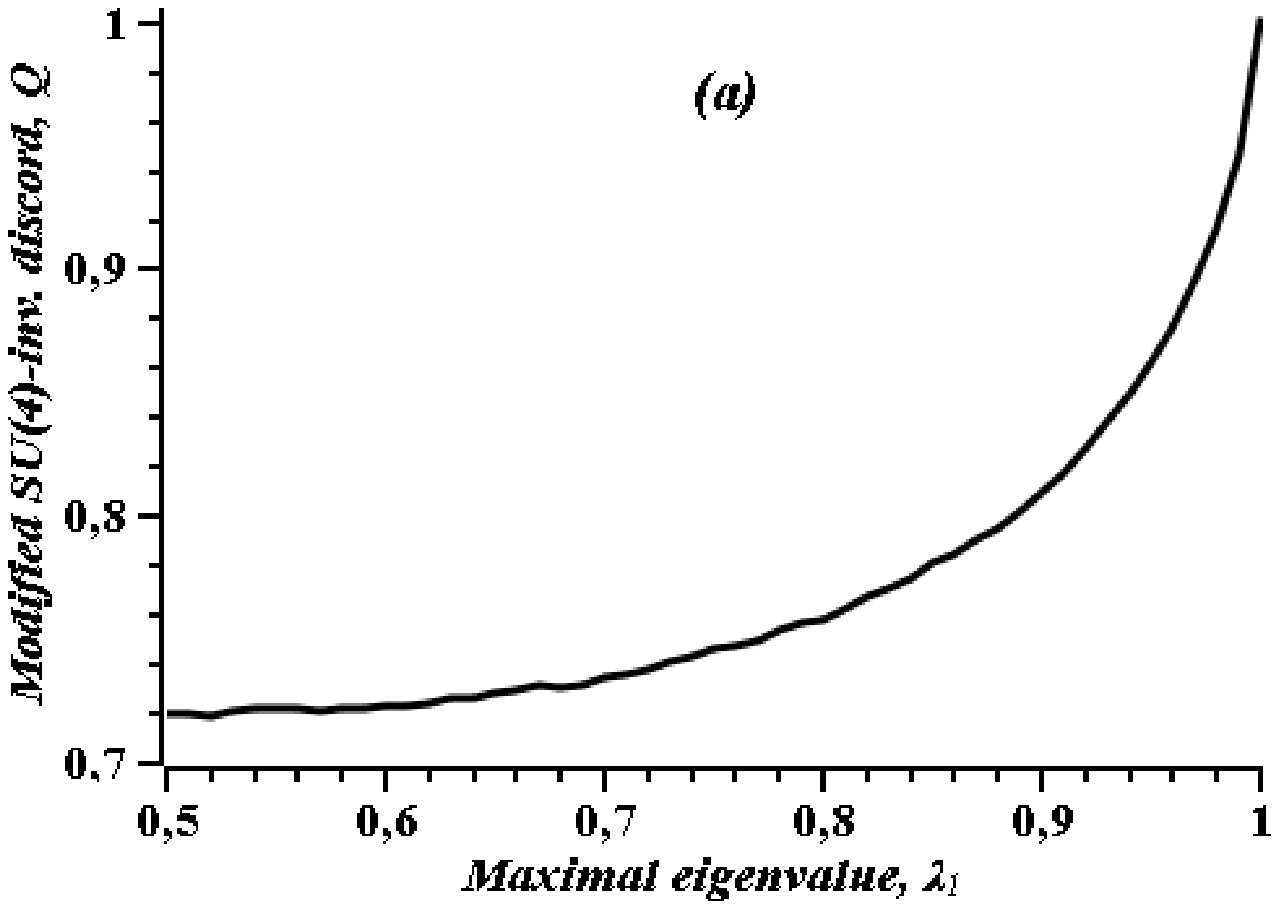
   ,scale=0.65,angle=0}
\epsfig{file=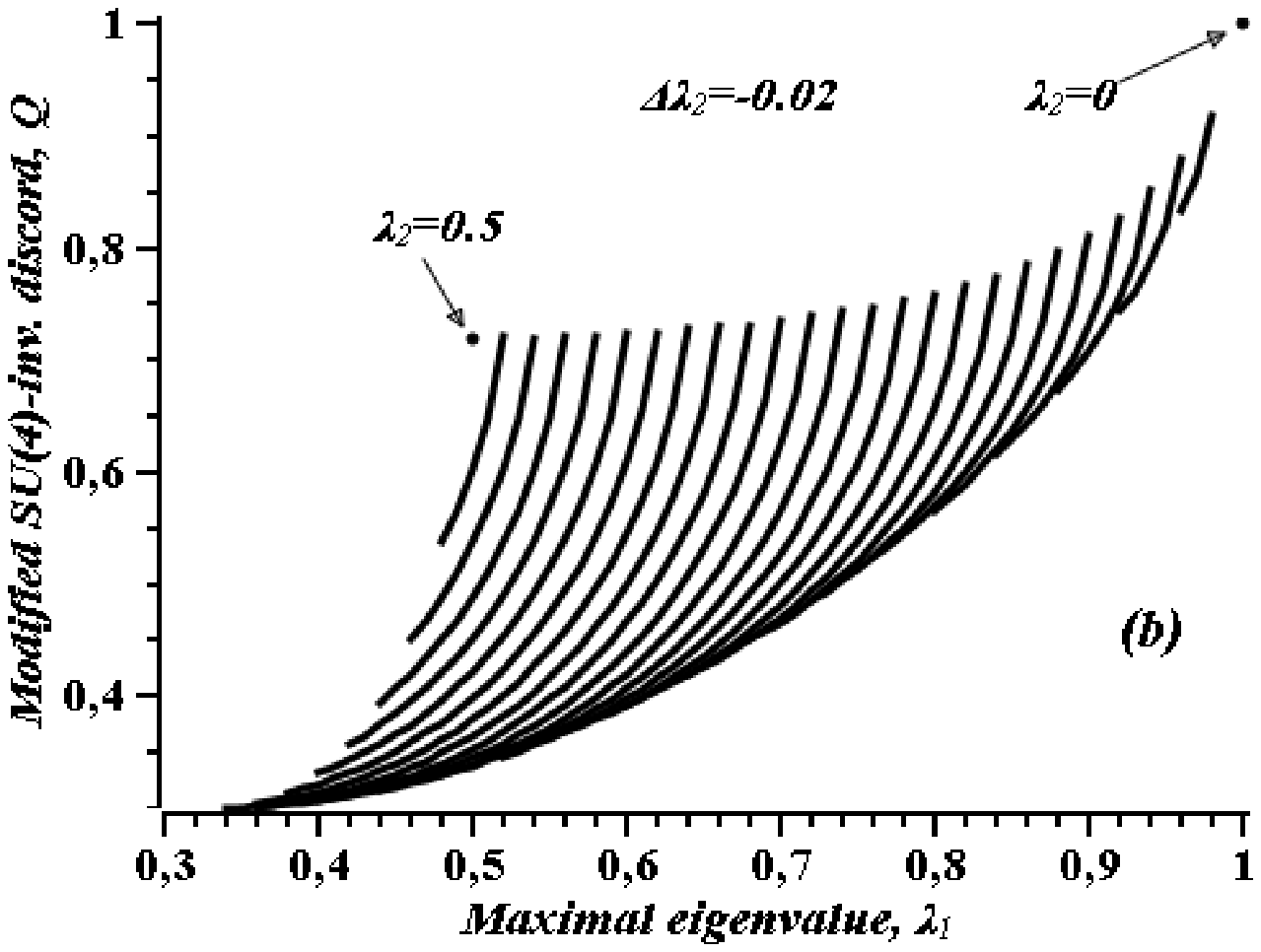
   ,scale=0.65,angle=0}
\epsfig{file=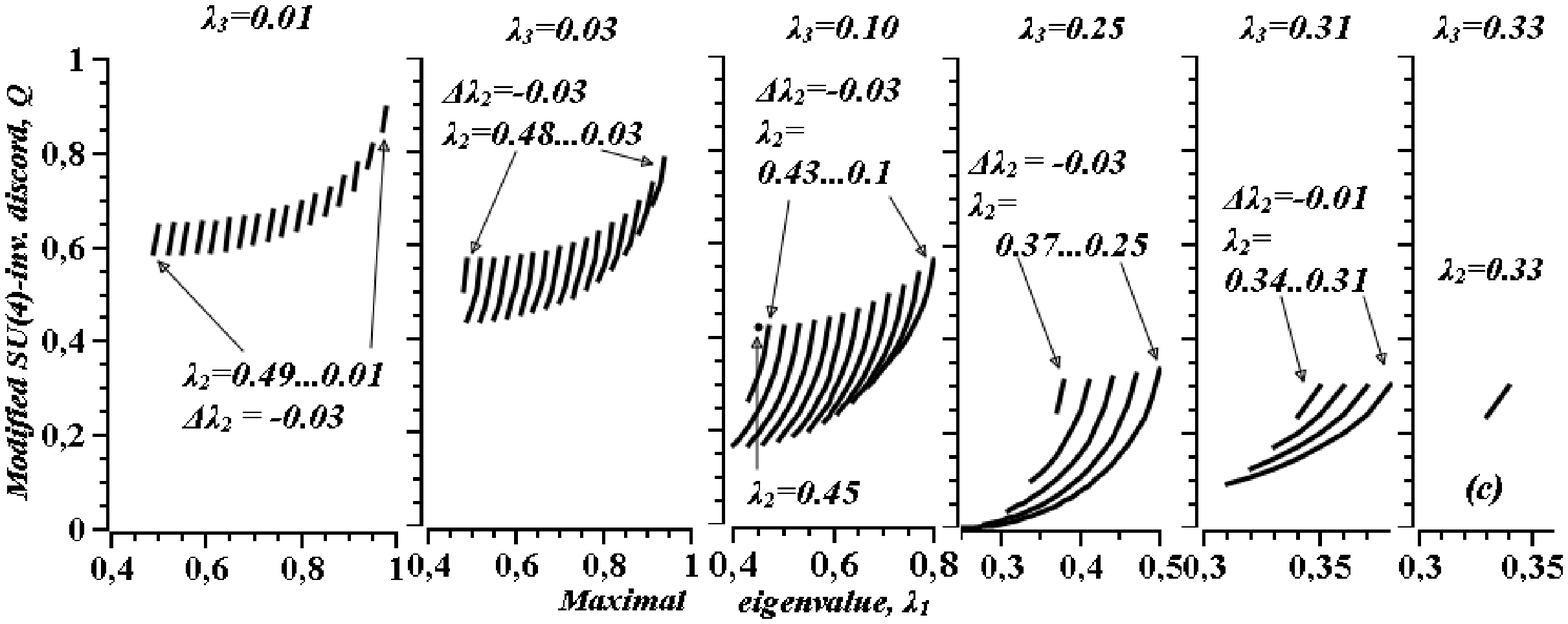
   ,scale=0.7,angle=0}
  \caption{The modified
$SU(4)$-invariant discord as a function of the maximal eigenvalue $\lambda_1$ for different numbers of nonzero eigenvalues;
($a$) at most two nonzero eigenvalues $\lambda_1$ and $\lambda_2$: $\lambda_2=1-\lambda_1$, $\lambda_3=\lambda_4=0$; ${\cal{Q}}\gtrsim 0.72$;
($b$) at most three nonzero eigenvalues $\lambda_i$, $i=1,2,3$: $\lambda_3=1-\lambda_1-\lambda_2$, $\lambda_4=0$; ${\cal{Q}}\gtrsim 0.30$;
($c$) all eigenvalues may be nonzero: $\lambda_4=1-\lambda_1-\lambda_2-\lambda_3$; ${\cal{Q}}\ge 0$
}
  \label{Fig:12} \label{Fig:123} \label{Fig:1234} 
\end{figure*}

Next, consider the case of at most  three nonzero $\lambda_i$:  $\lambda_3=1-\lambda_1-\lambda_2$, $
\lambda_4=0$.
Thus, there are two independent parameters $\lambda_1$ and $\lambda_2$ in this case. The graphical representation of the function ${\cal{Q}}(\Lambda_2)\equiv {\cal{Q}}(\lambda_1,\lambda_2)$ is shown in Fig.\ref{Fig:123}$(b)$. Here ${\cal{Q}}\gtrsim 0.30$,
i.e. the modified  $SU(4)$-invariant discord may not be less then $\approx 0.30$ if there is one  zero eigenvalue.

Finally, the most general case of all nonzero eigenvalues is represented in Fig.\ref{Fig:1234}$(c)$. 
In this case ${\cal{Q}}(\Lambda_2)\equiv {\cal{Q}}(\lambda_1,\lambda_2,\lambda_3)\ge 0$. Note that the modified $SU(4)$-invariant discord (as well as the  $SU(4)$-invariant discord itself)  equals zero only in the case 
$\Lambda_2 = \frac{1}{4} I_4$, see Sec.\ref{Section:zero_d} and Fig.\ref{Fig:1234}$(c)$, the curve with $\lambda_2=\lambda_3=0.25$.  As opposite, there is a family of states where the usual discord is zero \cite{OZ}.
Fig.\ref{Fig:12} demonstrates that the modified $SU(4)$- invariant discord does not exceed unit (${\cal{Q}}\le 1$) with 
${\cal{Q}}(\Lambda_2^{pure})=1$. 
In general, we conclude that in order  to maximize the modified $SU(4)$-invariant discord, one has to build the density matrix with as  many as possible vanishing eigenvalues. In particular, the condition $(1-\lambda_1)\ll 1$ may serve as a sufficient condition for the large modified $SU(4)$-invariant discord.

\section{Bipartite unitary invariant discord in an $N$-qubit system}
\label{Section:Nparticles}
Let us introduce the unitary invariant  discord  for the $N$-qubit system, $N>2$  (or for the system of $N>2$  spin-1/2 particles). Similar to the two-qubit system, we take the basis of the eigenvectors of the density matrix arranged in the decreasing order of its eigenvalues as the original basis (i.e. the  basis corresponding to $\varphi=0$). 
The density matrix  is diagonal in this basis:
\begin{eqnarray}\label{NLam}
&&
\rho(0)=\Lambda_N={\mbox{diag}} \{\lambda_1,\dots,\lambda_{2^N}\},\;\;\;\lambda_{2^N}=1-\sum_{i=1}^{2^N-1}\lambda_i, \;\;
\\\nonumber
&&
0\le \lambda_i\le 1,\;\;\;\lambda_1\ge \lambda_2\ge \dots \ge \lambda_{2^N},
\end{eqnarray}
which is the $N$-qubit analogy of 
eq.(\ref{Lam}).
Next, using this basis, we consider any two subsystems  $A$ and $B$ with $N_A$ and $N_B$  spin-1/2 particles respectively, $N_A+N_B=N$. Now we may  introduce   the $SU(2^N)$-invariant bipartite discord $Q_{AB}$ and its modification ${\cal{Q}}_{AB}$ just using 
 formulas (\ref{QQ}-\ref{calQ})  with the replacement $\Lambda_2\to\Lambda_N$ and with the vector parameter $\varphi$ having the appropriate  dimension: $\varphi=(\varphi_1,\dots,\varphi_{M_N})$, $M_N=2^{2N}-1$.
The two-side projectors $\Pi_{ij}$ will be generalized for the multiqubit case as follows:
\begin{eqnarray}
\Pi_{ij}(\varkappa) =\Pi^A_{i }(\varkappa_A)
  \otimes \Pi^B_{j}(\varkappa_B) ,\;\;\;\varkappa=(\varkappa_A,\varkappa_B),
\end{eqnarray}
where $\varkappa_{A,B}$ are  vector parameters $\varkappa_A=\{\varkappa_{A1},\dots,\varkappa_{A(2^{2 N_A}-1)}\}$, $\varkappa_B=\{\varkappa_{B1},\dots,\varkappa_{B(2^{2 N_B}-1)}\}$ and 
\begin{eqnarray}
&&
\Pi^A_{i}(\varkappa_A) = V^A(\varkappa_A)
\Pi^A_{i }(0) (V^A)^+(\varkappa_A),\;\;\; V^A(\varkappa_A)\in SU(2^{N_A}),\;\;\;i=1,\dots,2^{N_A},\\\nonumber
&&
\Pi^B_{i}(\varkappa_B) = V^B(\varkappa_B)
\Pi^B_{i }(0) (V^B)^+(\varkappa_B),\;\;\; V^B(\varkappa_B)\in SU(2^{N_B}),\;\;\;i=1,\dots,2^{N_B}.
\end{eqnarray}
Each of the  projection operators $\Pi^{A,B}_{i }(0)$, $i=1,\dots,2^{N_{A,B}}$, picks up the $i$th diagonal element of the proper subsystem:
\begin{eqnarray}
&&
\Pi^A_{i }(0)=|n_{i,1}\dots n_{i,N_A}\rangle \langle n_{i,1}\dots n_{i,N_A}|,\;\;i=1,\dots,2^{N_A},\\\nonumber
&&
\Pi^B_{j }(0)=|n_{j,1}\dots n_{j,N_B}\rangle \langle n_{j,1}\dots n_{j,N_B}|,\;\;j=1,\dots,2^{N_B},
\end{eqnarray}
where $n_{i,j}$ equal either 0 or 1 and $|n_{i,1}\dots n_{i,N_A}\rangle$ and $|n_{j,1}\dots n_{j,N_B}\rangle$ are bases in the  Hilbert subspaces corresponding  to the particles $A$ and $B$.
The multiqubit generalization of the eigenvalue matrix $\Lambda_2^{pure}$ (see eq.(\ref{Lam_2_pure})) is following:
\begin{eqnarray}\label{Lam_pure}
\Lambda_N^{pure}={\mbox{diag}}(\underbrace{1,0,\dots,0,0}_{2^N}).
\end{eqnarray}
In eqs.(\ref{NLam}) and (\ref{Lam_pure}), the eigenvalue $\lambda_1$ corresponds to the zero eigenvectors 
$|\underbrace{0\dots 0}_{2^{N_A}}\rangle$ and $|\underbrace{0\dots 0}_{2^{N_B}}\rangle$.

It is obvious that  the unitary invariant  discord $Q_{AB}$ (as well as the modified unitary invariant discord  ${\cal{Q}}_{AB}$) written for  any particular choice of $A$ and $B$ does not capture all possible bipartite  quantum correlations in the system of $N$ particles.
To take into account all possible bipartite quantum correlations we must consider the invariant discords of all possible  bipartite decompositions. Namely, we suggest the following 
definitions of the bipartite  unitary invariant discord and the  modified bipartite unitary invariant discord of the $N$-qubit quantum system:
\begin{eqnarray}\label{QN}
{{Q}}(\Lambda_N)=\min_{A,B} \;{{Q}}_{AB}(\Lambda_N),\;\;\;{\cal{Q}}(\Lambda_N)=\min_{A,B} \;{\cal{Q}}_{AB}(\Lambda_N),
\end{eqnarray}
where minimization is taken over all possible bipartite decompositions of the quantum system $A\cup B$. 
Considering the system of  two spin-1/2 particles we see that the minimization disappears from eqs.(\ref{QN})  so that  we obtain eqs.(\ref{QAB}).
Finally, let us note that the definition of the (modified) bipartite  unitary invariant discord  given by  eqs. (\ref{QN})  is similar to the characteristics of the multipartite entanglement  introduced in ref. \cite{Z2}: both use the minimization over all possible bipartite decompositions of a quantum system.

{{ Discords ${{Q}}(\Lambda_N)$ and ${\cal{Q}}(\Lambda_N)$ have the same four properties as respectively 
${{Q}}(\Lambda_2)$ and ${\cal{Q}}(\Lambda_2)$ (see Sec.\ref{Section:SU4}) with replacements
\begin{eqnarray}\label{repl}
\Lambda_2\to\Lambda_N,\;\;\; SU(4) \to SU(2^N).
\end{eqnarray}
.}

\subsection{The bipartite unitary invariant discord of the  reduced $N$-qubit system}

Now we turn to the discord between subsystems $A$ and $B$ consisting of $N_A$ and $N_B $ particles with $N_A+N_B < N$. As usual, the first step is the proper reduction of the original density matrix with respect to the subsystem $C$ consisting of  $N_C=N-N_A-N_B$ particles. However,  considering the unitary invariant discord, we are not forced to perform  the reduction 
in the system of original physical particles unlike the usual discord. Instead of physical particles, 
we are 
 free to choose a proper set of the virtual particles representing the given quantum system and perform  the reduction inside of this set. As it was written in the Introduction, the choice of the virtual particles is equivalent to the choice of the proper transformation $U(\varphi_1)\in  SU(2^N)$, which transforms the original basis  (which is the basis of the eigenvectors of $\rho$ arranged in  the decreasing order of  eigenvalues) into the multiplicative basis of the appropriate virtual particles, where the density matrix reads:
 $\rho(\varphi^1)= U(\varphi^1)\Lambda_N U^+(\varphi^1)$. 
Now we   reduce  $\rho(\varphi^1)$ with respect to the appropriate subsystem  $C$  and calculate  the  unitary invariant discord ${{Q}}_{C}$ 
using formulas of Sec.\ref{Section:Nparticles} with   the reduced density matrix having dimension $2^{N_A+N_B}$. Here ${{Q}}_{C}$ means the bipartite unitary invariant discord of the  quantum system reduced with respect to  the subsystem $C$.

There is a natural question whether the calculation of the discord in the reduced quantum system is a valuable procedure. 
In fact, the unitary invariant discord ${{Q}}$ given by the first eq.(\ref{QN}) takes into account  all bipartite correlations in a quantum system. 
Nevertheless, the unitary invariant discord in a reduced quantum system may be useful, for instance, 
in the following two cases.
\begin{enumerate}
\item
We are interested in the discord in the subsystem of $N_A+N_B<N$ original particles. 
In this case we consider the density matrix of quantum system in the multiplicative basis corresponding to the original particles and reduce it with respect to the subsystem $C$.
\item
As far as the reduced density matrix has lower dimension, the calculation of the bipartite unitary invariant  discord  for this matrix is simpler. This fact may be used to obtain an information about the overall quantum correlations in a large quantum system having the bipartite unitary invariant discords of the appropriate reduced subsystems, as it is done in the Sec.\ref{Section:example1}.
\end{enumerate}

\subsubsection{bi-particle unitary invariant discord of an $N$-qubit spin-1/2 system in a pure state}
\label{Section:example1}
We consider the case when the calculation of the unitary invariant  quantum discord in the reduced quantum system is helphull to demonstrate the presence of strong bipartite quantum correlations. Let us consider the single quantum state transfer along a spin chain \cite{Bose,CDEL,KF,KS,GKMT} and turn to the problem of relation between  the single quantum state transfer and  the quantum correlations. There are articles  where the relation between the state transfer probability and the entanglement is studied \cite{GMT,DFZ,DZ}. It is demonstrated that the perfect end-to-end state transfer probability and strong entanglement are not necessary related with each other. 
It  has also been shown  recently that one can arrange either the high probability quantum state transfer or the strong entanglement between any two particles in the spin-1/2 chain of $N$ nodes governed by the $H_{XYZ}$ Hamiltonian with the inhomogeneous magnetic field \cite{DZ} (similar numerical experiments with discord have not been evaluated yet). 
This fact may be taken as  an  evidence of the quantum correlations which  supplement the quantum state transfers  in the above systems. Of course, these correlations may be caught by the unitary invariant discord (\ref{QN}). However, calculations are very complicated because of the numerous  parameters in $SU(2^N)$. 
For this reasong we use the advantage of the reduced density matrix. Namely, we show that an arbitrary system of  $N$ spin-1/2 particles in a pure state may be viewed as a system of   $N$ virtual spin-1/2 particles with  
\begin{eqnarray}\label{Qij}
Q_{ij;rest}={\cal{Q}}_{ij;rest}=1
\end{eqnarray}
 for any two virtual particles $i$ and $j$. Here $ Q_{ij;rest}$ and ${\cal{Q}}_{ij;rest}$ mean the $SU(4)$-invariant discord and the modified  $SU(4)$-invariant discord between virtual particles $i$ and $j$ in a system of  $N$ virtual spin-1/2 particles  reduced with respect to all other virtual particles. The above  system  of virtual particles corresponds to the basis of ordered eigenvectors of the density matrix, i.e. $\rho(0)=\Lambda_N^{pure}$, see eq.(\ref{Lam_pure}). 

First of all remember that the only nonzero eigenvalue $\lambda_1=1$ corresponds to the zero eigenvectors $|\underbrace{0\dots 0}_{2^{N_A}}\rangle$ and $|\underbrace{0\dots 0}_{2^{N_B}}\rangle$, as it is pointed after eq.(\ref{Lam_pure}).
It is simple to show that the reduced density matrix with respect to any $(N-2)$ virtual particles reads
\begin{eqnarray}\label{Lam_red}
\rho^{red}\equiv \Lambda_2={\mbox{diag}}(1,0,0,0),
\end{eqnarray}
{ which is the density matrix of a pure state (\ref{Lam_2_pure}).
Thus  $ Q_{ij;rest}(\Lambda_2)={\cal{Q}}_{ij;rest}(\Lambda_2)=1$ for any two particles $i$ and $j$, and eqs.(\ref{Qij}) are valid.}
This allows us to conclude that the bipartite quantum correlations in the whole spin system are big. 

This conclusion is  applicable to 
the process of a single quantum state transfer along the spin-1/2 chain of $N$ nodes since the initial state is a pure one in this process. 
Thus we conclude that  a single  quantum state transfer along a spin chain  is directly related with strong quantum correlations  in this spin chain.

\section{Geometric measure of  a bipartite unitary invariant discord}
\label{Section:geometric}
{{ A geometric measure of either quantum 
entanglement or quantum discord has been introduced as a characteristics  showing how far is a given state from that having zero entanglement or  discord. It is defined by the formula \cite{HHH,WG,DVB,LF}}
\begin{eqnarray}\label{GG}
{{q}}^G_{AB}(\Lambda_N)=\int d\Omega(\varphi) \inf_{\chi}||\rho(\varphi) -\chi||^2,
\end{eqnarray}
where $\chi$  is the complete set of density matrices having zero measure of quantum correlations (either entanglement or discord). Here  $||\rho||^2 = {\mbox{Tr}}(\rho^2)$. In our case, $\chi$ is the complete  set of  matrices having such  ordered matrices of eigenvalues $\Lambda^{\chi}_N$  that ${{Q}}(\Lambda^{\chi}_N)=0$.  Thus, first of all, one has to define the appropriate set  of matrices $\chi$. 

\subsection{Quantum state with zero bipartite unitary invariant discord }
\label{Section:zero_d}

Let us show that the bipartite unitary invariant discord introduced by the first formula (\ref{QN}) is zero only if all eigenvalues of the density matrix  equal each other, i.e. 
\begin{eqnarray}\label{Lam_I}
\chi\equiv \Lambda_N=\frac{1}{2^N} I_{2^N}.
\end{eqnarray}
 First of all, remember that the usual discord is zero for the density matrix $\rho(\varphi_0)$ 
 which may be diagonalized by the local unitary transformations $V^A \in SU(2^{N_A})$ and $V^B \in SU(2^{N_B})$  \cite{OZ,X}:
\begin{eqnarray}\label{VV}
\rho(\varphi_0)=V^A(\varkappa_A(\varphi_0))\otimes V^B(\varkappa_B(\varphi_0)) \tilde\Lambda_N (V^A)^+(\varkappa_A(\varphi_0))\otimes (V^B)^+(\varkappa_B(\varphi_0)),
\end{eqnarray}
where $\tilde \Lambda_N$ is a diagonal matrix of eigenvalues of $\rho(\varphi_0)$. Note that the eigenvalues in $\tilde \Lambda_N$ are not necessary ordered, but they may be ordered by the appropriate orthogonal transformation $\tilde O\in SO(2^N)$:
\begin{eqnarray}\label{O}
\Lambda_N=\tilde O^+ \tilde \Lambda_N \tilde O.
\end{eqnarray} 
Here $\Lambda_N$ is an ordered matrix of eigenvalues (\ref{NLam}).
Regarding the unitary invariant discord, it may be zero only if eq.(\ref{VV}) is valid for any $\varphi_0$, which is a consequence of  the nonnegativity of discord.  Let us find the set of such matrices $\Lambda_N$ that satisfy this condition. 
As usual, we start with the basis of eigenvectors of $\rho$ which are arranged in the decreasing order of eigenvalues, i.e. $\rho(0)=\Lambda_N$ and $\rho(\varphi)=U(\varphi)\Lambda_N U^+(\varphi)$.  Then the unitary invariant  discord ${{Q}}(\Lambda_N)$ is zero if, for any $\varphi$,  there is such local transformation $V^A(\varkappa_A(\varphi))\in SU(2^{N_A})$ and $V^B(\varkappa_B(\varphi))\in SU(2^{N_B})$ that 
\begin{eqnarray}\label{tL}
V^A(\varkappa_A(\varphi))\otimes V^B(\varkappa_B(\varphi)) U(\varphi)  \Lambda_N U^+(\varphi) (V^A)^+(\varkappa_A(\varphi))\otimes (V^B)^+(\varkappa_B(\varphi)) = \hat \Lambda_N,
\end{eqnarray}
where $\hat \Lambda_N$ consists of the eigenvalues of $\rho(0)$ and, consequently, there is an orthogonal transformation $O\in SO(2^N) $ such that 
$\Lambda_N=O^+ \hat \Lambda_N O$.
As a consequence, eq.(\ref{tL}) may be written in the following form:
\begin{eqnarray}\label{LL}
[ V^A(\varkappa_A(\varphi))\otimes V^B(\varkappa_B(\varphi))U(\varphi)O^+  ,\hat \Lambda_N ]=0.
\end{eqnarray}
Since $ U(\varphi)\in SU(2^N)$, then the matrix  $V^A(\varkappa_A(\varphi))\otimes V^B(\varkappa_B(\varphi))U(\varphi)O^+ $ may not be diagonal for any $\varphi$. Thus, in order to satisfy condition (\ref{LL}), one has to assume that 
$\hat\Lambda_N$ is proportional to the identity matrix, i.e.
\begin{eqnarray}
\hat \Lambda_N=\frac{1}{2^N} I_{2^N}
\end{eqnarray}
{ independently on the particular choice of subsystems $A$ and $B$.}
This means that the state with zero unitary invariant discord is the state having all equal eigenvalues, i.e. 
eq.(\ref{Lam_I}) is valid. Then the manifold of matrices $\chi$ (see eq.(\ref{GG})) with fixed $N$ consists of the single element $\chi=\frac{1}{2^N} I_{2^N}$. This conclusion allows one to simplify eq.(\ref{GG}).

{{ In addition, one can show that the modified unitary invariant discord  (see the second  formula (\ref{QN})) is zero only for the density matrix (\ref{Lam_I}) similar to the unitary invariant discord. In fact, since $\bar q^{up}_{AB}\ge \bar q_{AB}$, then $\bar q^{up}_{AB}$ may be zero only if $\bar q_{AB}=0$. Thus, one has to show that $\bar q^{up}_{AB}$ is zero for the density matrix (\ref{Lam_I}). The later is true because, if $\rho(0) =\chi$, then $\rho(\varphi) =\chi$ as well, which means that $\rho^d(\varphi) = \rho(\varphi)$. Then eq.(\ref{barQup}) yields  
$\bar q^{up}_{AB} (\chi)=0$. Consequently, ${\cal{Q}}$ (defined by the second equation (\ref{QAB})) is zero only for the matrix $\chi$ given by eq.(\ref{Lam_I}). A direct consequence of this conclusion is that the geometric measure of modified unitary invariant discord coinsides with the geometric measure of unitary invariant discord.}

\subsection{Explicite formula for the geometric measure}
In accordance with Sec.\ref{Section:zero_d}, 
eq.(\ref{GG}) gets the following form:
\begin{eqnarray}\label{G}
&&
{{q}}^G_{AB}(\Lambda_N)=\int d\Omega(\varphi)||\rho(\varphi) -\frac{1}{2^N}I_{2^N}||^2=\\\nonumber
&&
\int d\Omega(\varphi)||U(\varphi)(\Lambda_N -\frac{1}{2^N}I_{2^N})U^+(\varphi)||^2=
||\Lambda_N -\frac{1}{2^N}I_{2^N}||^2.
\end{eqnarray}
Thus
\begin{eqnarray}
{{q}}^G_{AB}(\Lambda_N)=\sum_{i=1}^{2^N} (\lambda_i-\frac{1}{2^N})^2=\sum_{i=1}^{2^N} \lambda_i^2 -\frac{1}{2^N},
\end{eqnarray}
where we take into account that $\sum_i\lambda_i=1$.
We see that the geometric measure of the unitary invariant discord does not require neither optimization no multiple integration. 
It is obvious that this measure takes the maximal value for the pure state, when there is a single  nonzero eigenvalue $\lambda_1=1$:
\begin{eqnarray}\label{Gmax}
{{q}}^G(\Lambda^{pure}_N)=1 -\frac{1}{2^N}.
\end{eqnarray}
In particular, 
eq.(\ref{Gmax}) with $N=2$ yields ${{q}}^G(\Lambda^{pure}_2)=3/4>q(\Lambda^{pure}_2)\approx 0.42$.
It is reasonable to normalize the geometric measure as follows
\begin{eqnarray}\label{normG}
{{Q}}^G(\Lambda_N)=\frac{{{q}}^G(\Lambda_N)}{{{q}}^G(\Lambda_N^{pure})}=
\frac{2^N\sum_{i=1}^{2^N} \lambda_i^2 -1}{2^N-1},
\end{eqnarray}
thus ${{Q}}^G(\Lambda_N)\le 1$.  The quantity ${{Q}}^G(\Lambda_N)$ is referred to  as the normalized geometric measure of the unitary invariant discord (or normalized geometric measure  for the sake of brevity). {It possesses the same properties as ${{Q}}(\Lambda_N)$, see list of four properties in Sec.\ref{Section:SU4} with replacements (\ref{repl}).}
It is remarkable that  formula (\ref{normG}) remains correct for an arbitrary spin-1/2 quantum system. 

\section{Example: modified $SU(4)$-invariant discord  and  normalized geometric measure  of  thermal equilibrium states}

 \label{Section:Therm}
As an example, let us compare the  modified $SU(4)$-invariant discord with  the normalized geometric measure of  the system of  two spin-1/2 particles in the thermal equilibrium 
state \cite{QJH,WR,FZ}

\begin{eqnarray}
\rho=\frac{e^{-\frac{H}{kT}}}{Z},\;\;\; Z={\mbox{Tr}} \{e^{-\frac{H}{kT}}\}
\end{eqnarray}
 with different Hamiltonians $H$:
\begin{eqnarray}
\label{Heis}
H_{Heis}&=& { -D} (I_{1,x} I_{2,x}+I_{1,y} I_{2,y}+I_{1,z} I_{2,z} ) ,
\\
\label{XXZ}
H_{XYZ}&=&-D (I_{1,x} I_{2,x}+I_{1,y} I_{2,y} - 2 I_{1,z} I_{2,z}),\\\label{XY}
H_{XY}&=&-{D} (I_{1,x} I_{2,x}+I_{1,y} I_{2,y} ) .
\end{eqnarray}
Here $D$ (either positive or negative) is the constant of either dipole-dipole or exchange interaction, $T$ is the  temperature, $k$ is the Boltsman constant. 
In order to calculate the modified unitary invariant discord one needs the eigenvalues of Hamiltonians:
\begin{eqnarray}\label{Ham}
&&\lambda_{H_{Heis}}=\{-\frac{D}{4},-\frac{D}{4},-\frac{D}{4},\frac{3D}{4}\}
,\\\nonumber
&&\lambda_{H_{XYZ}}=\{0,-D,\frac{D}{2},\frac{D}{2}\},\\\nonumber
&&\lambda_{H_{XY}}=\{0,0,-\frac{D}{2},\frac{D}{2}\}.
\end{eqnarray}
The  normalized geometric measures of the thermal equilibrium states with Hamiltonians (\ref{Heis}-\ref{XXZ}) may be readily calculated using eq.(\ref{normG})  with  eigenvalues $\lambda_i$ corresponding to the eigenvalues of Hamiltonians (\ref{Ham}):
\begin{eqnarray}\label{GG2}
&&
Q^G_{Heis}=\left\{
\begin{array}{ll}\displaystyle
\left(\frac{e^{\beta}-1}{3 e^{\beta}+1}
\right)^2,& D>0\cr\displaystyle
\left(\frac{e^{\beta}-1}{ e^{\beta}+3}
\right)^2,& D<0
\end{array}
\right.,\\\nonumber
&&
Q^G_{XYZ}=\left\{
\begin{array}{ll}\displaystyle
\frac{3 e^{3\beta} -2 e^{2 \beta} - 4 e^{3\beta/2}+3e^\beta -4 e^{\beta/2} + 4}
{3(e^{3 \beta/2}+e^{\beta/2}+2)^2}
,& D>0\cr\displaystyle
\frac{4 e^{3\beta} -4 e^{5 \beta/2} +3 e^{2 \beta} - 4 e^{3\beta/2}-2e^\beta  + 3}
{3(2 e^{3 \beta/2}+e^{\beta}+1)^2}
,& D<0\end{array}
\right.,\\\nonumber
&&
Q^G_{XY}=
\frac{1}{3}
({\mbox{sech}}^4(\beta/4)-4 {\mbox{sech}}^2(\beta/4)+3)
, \;\;\;\forall \; D,
\end{eqnarray}
where $\beta=D/(kT)$. 
However, the modified unitary invariant discord itself may not be calculated analytically. Results of numerical calculations are represented in Fig.\ref{Fig:th} (thick lines), where the normalized  geometrical measures (\ref{GG2}) are shown for comparison (thin lines). 
\begin{figure*}
   \epsfig{file=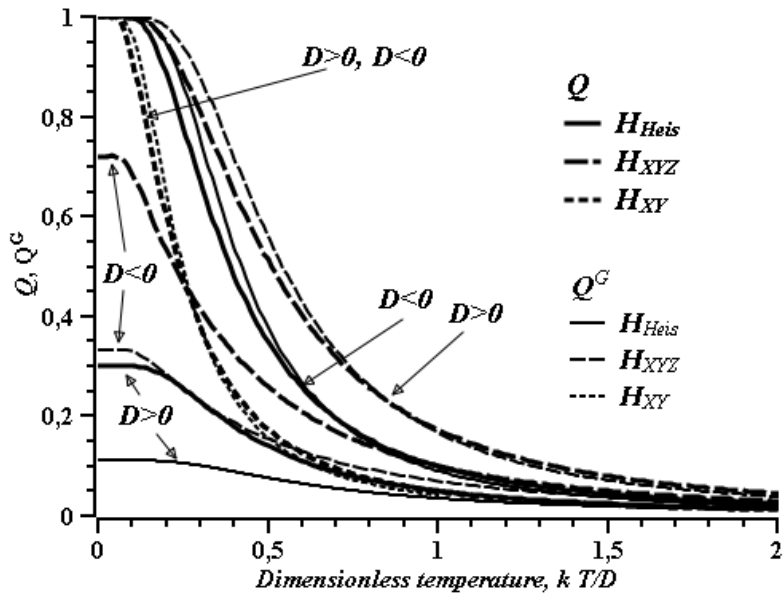
   ,scale=0.8,angle=0}
  \caption{The thermal equlibrium state of two spin-1/2 particles: the comparison of the modified $SU(4)$-invariant  discord  ${\cal{Q}}$ with the normalized geometric measure $Q^{G}$  for different Hamiltonians. Graph corresponding to the Hamiltonian $H_{XY}$  does  not depend on the sign of $D$ (there is a single thick and thin dashed lines) }
  \label{Fig:th} 
\end{figure*}

The different values of the modified  unitary invariant  discords of the ground states ($T=0$) are well explained by the  Hamiltonian eigenvalues (\ref{Ham}). 
Since the ground state corresponds to the minimal eigenvalue(s),
it   is a pure state in most of the considered cases such as
$H_{XY}$ ($\forall D\neq 0$), $H_{dz}$ ($D>0$) and $H_{Heis}$ ($D<0$), which correspond to ${\cal{Q}}={{Q}}^G=1$.
 The ground state has the degeneration  degree two in the case of the Hamiltonian 
 $H_{dz}$ with $D<0$, when the nonzero eigenvalues of the density matrix are $\lambda_1=\lambda_2=1/2$ and  ${\cal{Q}} \approx 0.72$, ${{Q}}^G =1/3$, so that ${\cal{Q}}-{{Q}}^G \approx 0.39$. The  ground state has the
degeneration  degree three in the case of Hamiltonian 
 $H_{Heis}$ with $D>0$, when the nonzero eigenvalues of the density matrix are 
$\lambda_1=\lambda_2=\lambda_3=1/3$ and ${\cal{Q}} \approx 0.30$, ${{Q}}^G =1/9$ so that ${\cal{Q}}-{{Q}}^G \approx 0.19$. 
The calculated values of ${\cal{Q}}$ of ground states are confirmed by  Fig.\ref{Fig:12}. 
We see that the difference  between ${\cal{Q}}$ and ${{Q}}^G$ is significant for the degenerate ground states.



\section{Conclusions}
\label{Section:conclusions}
We propose a $SU(2^N)$-invariant measure of bipartite quantum correlations in a system of  $N$ spin-1/2 particles. The feature of this measure is  that it takes into account not only the quantum correlations  between the original physical particles but also the quantum correlations  between all virtual particles, which  may  also be responsible for the advantages of quantum devices. 
The calculation of the unitary invariant discord seemed out to be a complicated computational problem involving both the multiparameter optimization and integration over the multidimensional parameter space of $SU(2^N)$ group. This forces us to
introduce the  modified   unitary invariant discord which requires only multidimensional integration without optimization and consequently may be  calculated simpler. The case $N=2$ is studied in more details. In particular, it is obtained that the maximal value of the modified $SU(4)$-invariant discord corresponds to a pure state. Moreover, considering a pure state of a multiqubit system we obtain that there is such complete set of $N$ virtual particles that the  unitary invariant discord  between any two  particles from this set achieves unit. This example is closely related with the  single quantum state transfer along the spin-1/2 chain as an evidence of strong quantum correlations which assist this process. This fact confirms the importance of the quantum correlations between the virtual particles in the quantum information processes.

Next, we introduce the  normalized geometric measure of unitary invariant discord { which coinsides with the  normalized geometric measure of modified unitary invariant discord}. Unlike the unitary invariant discord itself, its
geometric measure has simple analitical representation in terms of eigenvalues of the density matrix and, as a consequence, neither optimization nor multidimensional  integration is required for its calculation. 

{ The unitray invariant discord $Q$, the modified discord ${\cal{Q}}$ and the geometric measure $Q^G$ possess the same list of four properties given in Sec.\ref{Section:SU4} with replacements (\ref{repl}). In addition, it is shown that ${\cal{Q}}(\Lambda_2) \le 1$ and $Q^G(\Lambda_N)\le 1$. 

The calculation of the unitary invariant discord is a complicated procedure involving both optimization and multiple integration. The calculation of the modified unitary invariant discord is simpler, because it involves only multiple integration. But both of the above discords require numerical calculation. However, the calculation of the geometric measure is quite simple and may be performed analytically for the states of system with any number of spin-1/2 particles using eq.(\ref{normG}). This property of the geometric measure allows one to use it as a witness of quantum correlations which is especially important because the calculation of the usual  discord using the algorithm described in  \cite{OZ} is a  complicated procedure and can be done only for particular states \cite{L,ARA,X}. Calculating the geometric measure using eq.(\ref{normG})  we readily answer the question whether the given system possesses quantum correlations.}

The modified unitary invariant discords and the normalized geometric measures of both the  ground states and the thermal equilibrium states for the systems governed by different Hamiltonians are compared. 
 Fig.\ref{Fig:th} shows that the modified $SU(4)$-invariant  discords as well as the normalized geometric measures of $SU(4)$-invariant discord of the ground states ($T=0$) decrease with the increase in the degeneration degree of the ground states.

Altough the unitary invariant discord is introduced for  the $N$-qubit system, it may be straightforwardly  generalized  for an arbitrary quantum system. The same  is valid for the geometric measure. 

The author thanks Professor E.B.Fel'dman for usefull discusions and advices.
This work is supported 
by the Program of the Presidium of RAS 
No.18 "Development of methods of obtaining chemical compounds and creation of new
materials".

\section{Appendix}
The  basis for the Lie algebra of  $SU(4)$ \cite{GM}:
\begin{eqnarray}\label{gamma}
&&
\gamma_1=\left[
\begin{array}{cccc}
         0& 1& 0& 0\cr
         1& 0& 0& 0\cr
         0& 0& 0& 0\cr
         0& 0& 0& 0  
\end{array}
\right],\;\;
\gamma_2=\left[
\begin{array}{cccc}
         0&-i& 0& 0\cr
         i& 0& 0& 0\cr
         0& 0& 0& 0\cr
         0& 0& 0& 0  
\end{array}
\right],\;\;
\gamma_3=\left[
\begin{array}{cccc}
         1& 0& 0& 0\cr
         0&-1& 0& 0\cr
         0& 0& 0& 0\cr
         0& 0& 0& 0  
\end{array}
\right],\\\nonumber
&&
\gamma_4=\left[
\begin{array}{cccc}
         0& 0& 1& 0\cr
         0& 0& 0& 0\cr
         1& 0& 0& 0\cr
         0& 0& 0& 0  
\end{array}
\right],\;\;
\gamma_5=\left[
\begin{array}{cccc}
         0& 0&-i& 0\cr
         0& 0& 0& 0\cr
         i& 0& 0& 0\cr
         0& 0& 0& 0  
\end{array}
\right],\;\;
\gamma_6=\left[
\begin{array}{cccc}
         0& 0& 0& 0\cr
         0& 0& 1& 0\cr
         0& 1& 0& 0\cr
         0& \;\;0& 0& 0  
\end{array}
\right],\\\nonumber
&&
\gamma_7=\left[
\begin{array}{cccc}
         0& 0& 0& 0\cr
         0& 0&-i& 0\cr
         0& i& 0& 0\cr
         0& 0& 0& 0  
\end{array}
\right],\;\;
\gamma_8=\frac{1}{\sqrt{3}}\left[
\begin{array}{cccc}
         1& 0& 0& 0\cr
         0& 1& 0& 0\cr
         0& 0&-2& 0\cr
         0& 0& 0& 0  
\end{array}
\right],\;\;
\gamma_9=\left[
\begin{array}{cccc}
         0& 0& 0& 1\cr
         0& 0& 0& 0\cr
         0& 0& 0& 0\cr
         1& 0& 0& 0  
\end{array}
\right],\\\nonumber
&&
\gamma_{10}=\left[
\begin{array}{cccc}
         0& 0& 0&-i\cr
         0& 0& 0& 0\cr
         0& 0& 0& 0\cr
         i& 0& 0& 0  
\end{array}
\right],\;\;
\gamma_{11}=\left[
\begin{array}{cccc}
         0& 0& 0& 0\cr
         0& 0& 0& 1\cr
         0& 0& 0& 0\cr
         0& 1& 0& 0  
\end{array}
\right],\;\;
\gamma_{12}=\left[
\begin{array}{cccc}
         0& 0& 0& 0\cr
         0& 0& 0&-i\cr
         0& 0& 0& 0\cr
         0& i& 0& 0  
\end{array}
\right],\\\nonumber
&&
\gamma_{13}=\left[
\begin{array}{cccc}
         0& 0& 0& 0\cr
         0& 0& 0& 0\cr
         0& 0& 0& 1\cr
         0& 0& 1& 0  
\end{array}
\right],\;\;
\gamma_{14}=\left[
\begin{array}{cccc}
         0& 0& 0& 0\cr
         0& 0& 0& 0\cr
         0& 0& 0&-i\cr
         0& 0& i& 0  
\end{array}
\right],\;\;
\gamma_{15}=\frac{1}{\sqrt{6}}\left[
\begin{array}{cccc}
         1& 0& 0& 0\cr
         0& 1& 0& 0\cr
         0& 0& 1& 0\cr
         0& 0& 0&-3  
\end{array}
\right].
\end{eqnarray}


\begin{thebibliography}{99}



\bibitem{W}
R.F.Werner, Phys.Rev.A {\bf  40}, 4277 (1989)


\bibitem{HW}
S.Hill and W.K.Wootters, Phys. Rev. Lett. {\bf  78}, 5022 (1997)



\bibitem{P}
A.Peres, Phys. Rev. Lett. {\bf  77}, 1413 (1996)


\bibitem{AFOV}
L.Amico, R.Fazio, A.Osterloh and V.Ventral, Rev. Mod. Phys. {\bf  80}, 517 
(2008)

\bibitem{HHHH}
R.Horodecki,
P.Horodecki, M.Horodecki and K.Horodecki, Rev. Mod. Phys. {\bf  81}, 865  (2009)

{
\bibitem{BDFMRSSW}
C.H.Bennett, D.P.DiVincenzo, C.A.Fuchs, T.Mor, E.Rains, P.W.Shor, J.A.Smolin, and W.K.Wootters,
    Phys. Rev. A {\bf  59}, 1070 (1999).

\bibitem{HHHOSSS}
M.Horodecki, P.Horodecki, R.Horodecki, J.Oppenheim, A.Sen, U.Sen, and B.Synak-Radtke, Phys.Rev.A {\bf  71}, 062307 (2005).

\bibitem{NC}
 J.Niset and N. J.Cerf, Phys.Rev.A {\bf  74}, 052103 (2006)
}

\bibitem{M}
 D.A.Meyer, Phys. Rev. Lett. {\bf  85}, 2014 (2000).

\bibitem{DFC}
 A.Datta, S.T.Flammia, and C.M.Caves, Phys. Rev. A {\bf  72}, 042316 (2005); 

\bibitem{DV}
A.Datta and G.Vidal,   Phys. Rev. A {\bf  75}, 042310 (2007); 

\bibitem{DSC}
A.Datta, A.Shaji, and C.M.Caves, Phys. Rev. Lett. {\bf  100}, 050502 (2008).

\bibitem{LBAW}
B.P.Lanyon, M.Barbieri, M.P.Almeida, and A.G.White, Phys. Rev. Lett. {\bf  101}, 200501 (2008).

\bibitem{OZ}
H.Ollivier and W.H.Zurek, Phys.Rev.Lett. {\bf  88}, 017901(2001)


\bibitem{L}
S.Luo, Phys.Rev.A {\bf  77}, 042303 (2008)

\bibitem{ARA}
M.Ali, A.R.P.Rau and G.Alber, Phys.Rev.A {\bf  81}, 042105

\bibitem{Z}
W.H.Zurek, Phys.Rev.D {\bf  24}, 1516 (1981)


\bibitem{LL}
N.Li and S.Luo, Phys.Rev.A {\bf  76}, 032327 (2007)

\bibitem{L2}
S.Luo,  Phys.Rev.A {\bf  77}, 022301 (2008) 

\bibitem{FZ}
E.B.Fel'dman and A.I.Zenchuk, JETP Letters, {\bf  93},  459 (2011)

\bibitem{X}
J.-W. Xu, arXiv:1101.3408 [quant-ph]

{ 
\bibitem{HHH}
M.Horodecki,
P.Horodecki and R.Horodecki, Phys.Lett. A {\bf  223}, 1  (1996)

\bibitem{WG}
T.C.Wei and P.M.Goldbart, Phys. Rev. A {\bf  68}, 042307 (2003)
}

\bibitem{DVB}
B.Dakic, V.Vedral and C.Brukner, Phys.Rev.Lett. {\bf  105}, 190502 (2010)

\bibitem{LF}
S.Luo and S.Fu, Phys.Rev.A {\bf  82}, 034302 (2010)

\bibitem{TBS}
T.Tilma, M.Byrd and E.C.G.Sudarshan, J.Phys.A:Math.Gen. {\bf  35}, 10445 (2002)


\bibitem{Z2}
A.I.Zenchuk, 	
arXiv:1005.5531v2 [quant-ph]


\bibitem{Bose}
S.Bose, Phys.Rev.Lett, {\bf  91}, 207901 (2003)

\bibitem{CDEL}
M.Christandl, N.Datta, A.Ekert and A.J.Landahl, Phys.Rev.Lett. {\bf  92}, 187902 (2004)

\bibitem{KF}
E.I.Kuznetsova and E.B.Fel'dman, J.Exp.Theor.Phys., {\bf  102}, 882 (2006)

\bibitem{KS}
P.Karbach and J.Stolze, Phys.Rev.A, {\bf  72}, 030301(R) (2005)

\bibitem{GKMT}
G.Gualdi, V.Kostak, I.Marzoli and P.Tombesi, Phys.Rev. A, {\bf  78}, 022325 (2008)

\bibitem{GMT}
G.Gualdi, I.Marzoli and P.Tombesi, New J.Phys. {\bf  11}, 063038 (2009)

\bibitem{DFZ}
S. I. Doronin, E. B. Fel'dman, and A. I. Zenchuk,
Phys.Rev.A {\bf  79},
042310 (2009) 

\bibitem{DZ}
S.I.Doronin, A.I.Zenchuk, 
 Phys.Rev.A {\bf  81}, 022321 (2010)  

\bibitem{QJH}
W.Qiong, L.Jie-Qiao and Z. Hao-Sheng, Chin.Phys.B, {\bf  19}, 100311 (2010)

\bibitem{WR}
T.Werlang and G.Rigolin, Phys.Rev.A, {\bf  81}, 044101 (2010)

\bibitem{GM}
W.Greiner and B.M\"uller: {\it Quantum Mechanics: Symmetries}, (Springer, Berlin 1989)

\end{thebibliography}
\end{document}